# Investigation of cosmic ray–cloud connections using MISR

Joshua Krissansen-Totton and Roger Davies

Department of Physics, The University of Auckland, Auckland, New Zealand

Numerous empirical studies have analyzed International Satellite Cloud Climatology Project (ISCCP) data and reached contradictory conclusions regarding the influence of solar-modulated galactic cosmic rays on cloud fraction and cloud properties. The Multiangle Imaging SpectroRadiometer (MISR) instrument on the Terra satellite has been in continuous operation for 13 years and thus provides an independent (and previously unutilized) cloud dataset to investigate purported solar–cloud links. Furthermore, unlike many previous solar–climate studies that report cloud fraction MISR measures albedo, which has clearer climatological relevance. Our long-term analysis of MISR data finds no statistically significant correlations between cosmic rays and global albedo or globally averaged cloud height, and no evidence for any regional or lagged correlations. Moreover, epoch superposition analysis of Forbush decreases reveals no detectable albedo response to cosmic ray decreases, thereby placing an upper limit on the possible influence of cosmic ray variations on global albedo of 0.0029 per 5% decrease. The implications for recent global warming are discussed.

## 1. Introduction

The analysis of paleoclimate records has uncovered numerous robust correlations between climate proxies and indicators of solar activity [*Bond et al.*, 2001; *Neff et al.*, 2001]. Although current knowledge of long-term variations in total solar irradiance (TSI) is incomplete, arguably these close correlations cannot be fully explained by changes in globally averaged TSI alone [*Kirkby*, 2007]. This suggests the existence of some indirect solar influence on Earth's climate that amplifies relatively small changes in TSI.

Two broad categories of mechanisms have been proposed to explain observed solar–climate connections: indirect irradiance mechanisms [e.g. *Haigh*, 1996; *Meehl et al.*, 2009] and solar-modulated cosmic ray (CR) mechanisms, the latter being the focus of this study. During times of high solar activity, the sun's intensified magnetic field sweeps away galactic CRs from within the solar system, thus reducing the CR flux recorded on earth and ensuring the CR flux and solar activity are negatively correlated [*Bazilevskaya et al.*, 2008]. It has been argued that solar-modulated variations in CRs may in turn alter cloud properties via modulation of the global electric circuit [*Tinsley*, 2008] or changes in CCN formation [*Marsh and Svensmark*, 2000]. The prediction of the latter hypothesis is that increasing the CR flux increases atmospheric ionization, which in turn leads to higher CCN populations and therefore higher albedos. Unfortunately it is difficult to distinguish between competing mechanisms from observations alone since all indicators of solar activity such as CRs, TSI, sunspot number and the UV flux are all closely correlated [*Gray et al.*, 2010]. However despite this causal ambiguity it is helpful to determine whether and to what extent solar–climate correlations exist as a stepping-stone to understanding potential mechanisms.

Various empirical studies have claimed strong correlations exist between galactic CRs and satellite-detected cloud cover on interannual timescales [e.g. *Svensmark and Friis-Christensen*, 1997; *Pallé and Butler*, 2000; *Marsh and Svensmark*, 2003]. From these correlations it has been argued that the radiative forcing due to galactic CR modulation of cloud fraction could potentially explain most of the global warming observed in the 20th century [*Svensmark*, 2007; *Rao*, 2011]. However the choice of data and methods of analysis in these correlational studies have been heavily criticized [e.g. *Jorgensen and Hansen*, 2000; *Damon and Laut*, 2004], and similar empirical analyses by *Kristjansson et al*. [2004] and *Sun and Bradley* [2002] have not supported a galactic CR–climate link.

An alternative approach for investigating links between solar activity and climate is the epoch superposition of Forbush decreases (see references below). Forbush decreases (Fds) are sudden decreases in the surface CR flux caused predominantly by coronal mass ejections. The decrease takes place over several hours whilst the recovery back to original levels may take several days. By superposing multiple Fds and comparing their composite to a cotemporal superposition of atmospheric data it is possible to test whether sudden decreases in CRs are correlated with an atmospheric response. Since Fds are comparable in magnitude to CR variation over the 11-year solar cycle, purported correlations between CRs and climate on interannual timescales may also be apparent during Fds. Furthermore the short timescales involved allow internal modes of variability to be excluded as competing explanations for any correlations or trends.

Numerous composite studies have investigated possible atmospheric responses to Fds with positive results [e.g. *Pudovkin and Veretenenko*, 1995; *Todd and Kniveton*, 2001, 2004; *Svensmark et al*., 2012]. However many similar studies have obtained results consistent with the null hypothesis [e.g. *Kristjansson et al*., 2008; *Sloan and Wolfendale*, 2008; *Calogovic et al*., 2010; *Laken and Calogovic*, 2011].

Part of the reason why definitive results have proven elusive is because the majority of existing satellite-based studies are not independent; with the exception of a small number of studies that utilized MODIS, almost all empirical solar–climate studies have relied on ISCCP. The over-reliance on a single dataset is problematic because there are artefacts within the ISCCP dataset that arguably make it unsuitable for long-term analysis [*Laken et al*., 2012 and references therein]. The MISR instrument on the Terra satellite has been in continuous operation for 13 years and thus provides an independent and well-calibrated cloud dataset for investigating both long-term and short-term connections between CRs and climate. To our knowledge this study is the first time MISR has been used to explore solar–climate links.

## 2. Data

CR data were obtained from the Neutron Monitor Database (NMDB). For the long-term analysis, CR time series with daily count rates from eight different monitoring stations were anomalized, scaled by their own variances, and averaged to produce a 13-year CR composite to represent the normalized global CR flux (see supplementary material for list of monitoring stations).

Albedo and cloud height data were obtained from the MISR dataset. MISR utilizes 9 multi-channel cameras positioned at different viewing angles; this allows cloud-top heights to be stereoscopically derived, and top-of-atmosphere albedos to be inferred by integrating bidirectional reflectance factors over all 9 zenith angles and modeling azimuthal angle dependency [*Diner et al.*, 1999]. Thus MISR's 'expansive' albedo measurements mimic what an albedometer would measure if placed 30km above the surface. By comparison, ISCCP albedo estimates involve modeling of both azimuthal and zenith directionality and only attempt to determine the albedo at the top of the local reflecting layer. Although azimuthal directionality must still be modeled in MISR's case, the main result is differential so biases due to modeling cancel out [*Diner et al.*, 1999]. We analyzed the entire MISR dataset of expansive albedo values and zero-wind reflecting layer reference altitudes values (cloud-top height) available from the level 2 processing for the time period April 2000 to February 2013. These data were initially summarized in 140km along-track by 380km across-track 'blocks'. There are 180 blocks per orbit and MISR completes 14.56 orbits per day.

Many of the solar–climate studies referenced above have focused on cloud fraction as opposed to albedos, which are the subject of this study. The advantage of using albedos to study solar–climate links is that variations in albedo have unambiguous climatological impacts, whereas variations in cloud fraction have no such direct climatological correspondence since they may be compensated for by changes in optical depth. Few studies have explored the possibility that CRs could influence cloud height, and so we also analyzed cloud heights in addition to albedos in case a connection has been overlooked.

## 3. Long-Term Analysis

### 3.1. Methodology

Global and regional monthly albedos were calculated by weighting each block by latitude and solar insolation, and the resulting time series were anomalized by subtracting monthly averages. Latitude weighted monthly cloud height anomalies were similarly calculated. Orbits were assumed to be independent for the purposes of calculating the standard error (SE) in albedo or cloud height anomalies for each time period. To calculate p-values for correlation coefficients, temporal autocorrelation was accounted for by calculating the effective sample size for each time series [*Chatfield*, 1996]. Unless otherwise stated $p<0.05$ (two-tailed) is taken to be a statistically significant correlation.

### 3.2. Results

Fig. 1 shows the global albedo anomaly (continuous red) and the normalized CR anomaly (dashed blue) for the 13 years of MISR's operation. The correlation coefficient between the two time series is -0.57. The high degree of autocorrelation in the CR time series means the effective sample size (3 months) is too small for the p-value to be meaningful, though it is still not significant. The negative correlation is largely attributable to the downward trend in global albedo due to sea-ice melt [*Davies*, 2013] and the apparent upward trend in CRs is due to the chosen start/end point in the 11-year solar cycle (see *Gray et al.* [2010] for recent CR flux increase in the context of previous solar cycles). If the two time series are detrended then the correlation coefficient increases to -0.10, the effective sample size increases to 5.3 months and the p-value of the correlation coefficient is 0.86.

To test the possibility that monthly variations in the CR flux are correlated with global albedo, the albedo time series was detrended and the 11-year solar cycle was removed by subtracting a 3-year running mean from the CR time series. The resulting time series are shown in the supplementary material (Fig. S1). The correlation coefficient between the two time series is 0.11, the effective sample size is 35.6 months and the p-value of the correlation is 0.52. When this analysis was repeated for cloud-top height there were also no statistically significant correlations with the CR flux on long timescales (not shown).

A lagged correlation was performed between global albedo and normalized CRs over a ±600 day period (Fig. S2). This was done for (a) neither albedos nor CRs detrended, (b) both time series detrended, and (c) albedos detrended and the 11-year solar cycle removed from the CR time series as described above. In all three cases the best correlation occurs for positive lags, which is the non-causal direction for CRs influencing albedos. We also observe that in all three cases there are no (causal) lag values for which the correlation is significant, which is consistent with the null hypothesis.

To test for regional CR–cloud connections, albedos were binned by latitude and longitude and the time series for each grid box was correlated with the (global) normalized CR anomaly. Fig. 2a shows the resulting correlation coefficients as a function of latitude and longitude, and Fig. 2b shows correlation coefficients as a function of latitude only (in both cases both time series were detrended). Evidently there is no obvious spatial structure to the correlations and the vast majority of correlation coefficients are weak, with 90% falling between -0.37 and 0.37. Note that if neither time series is detrended (Fig. S3) we observe strong negative correlations in the arctic attributable to the melt of sea-ice. The Eurasian regions of negative correlation in Fig. 2a have the wrong sign for mechanisms that amplify TSI changes, and they disappear when the 11-year solar cycle is removed (Fig. S4). Fig. 2a also shows a slight tendency for positive correlations near the South Pole, but high latitudes have zero retrievals in winter and the correlations disappear when the 11-year solar cycle is removed. In short there is no evidence for any regional connections between CRs and albedos.

## 4. Forbush Decrease Analysis

### 4.1. Methodology

The Kerguelen Neutron Monitor (R=1.14 GV) was used to represent the global CR flux for the Fd analysis instead of a normalized eight station composite. 14 Fd events were chosen for the composite analysis (see supplementary material for list of event dates). Events were excluded from the composite if secondary Fd of comparable magnitude were apparent within ±20 days of the decrease minima, or if a strong (i.e. >2% increase within 1 day) ground level enhancement (GLE) was present within ±20 days of the minima. GLEs are brief increases (~hours) in the CR flux that occur when solar CRs cross the path of the Earth. They can be a confounding factor in Fd analyses since they act to offset coincident Fd events. The 14 Fd events were averaged by aligning their minima, and the analysis period was chosen to be 53 days (21 days prior the minima plus 31 following the minima).

Given the dates for each Fd, albedo and cloud height data for these dates were extracted from the MISR level 2 dataset. The 14 albedo time series that correspond to the Fd events were aligned and superposed in the same way to produce an event-averaged albedo time series. This was done both globally and regionally. The albedo time series, cloud height time series and CR time series were anomalized by subtracting a 21-day running mean. This was done to remove intermediate to long timescale trends/variations that were unrelated to the influence of sudden CR decreases (see *Laken and Calogovic* [2013] for further discussion).

To evaluate the significance of anomalies within the composite time series, and of correlations between different time series, Monte Carlo (MC) simulations were run for all Fd analyses using 14-event composites selected at random from the 13 years of MISR data. The resulting distributions of anomalies and correlation coefficients were used to evaluate statistical significance. The MC methodology we implemented is described in detail by *Laken and Calogovic* [2013].

### 4.2. Results

Fig. 3a shows the superposed CR time series for all 14 Fd events. The SE in the CR composite is also shown in addition to confidence intervals based on 100,000 MC simulations ($p<0.05$ and $p<0.01$ two-tailed levels are plotted). There is a clear statistically significant variation in the CR flux during Fds with 40% of days showing significant anomalies at the $p<0.05$ level. Fig. 3b shows the cotemporal broadband global albedo anomaly averaged over the same 14 events. There are only four days with anomalies significant at the $p<0.05$ level, and zero days with anomalies significant at the $p<0.01$ level which is consistent with the null hypothesis for a time series with 53 independent data points. Of the four statistically significant anomalies one is prior to the CR decrease and therefore non-causal, two have the wrong sign for the CCN mechanism suggested by *Marsh and Svensmark* [2000] and all four are only one day in duration, which is indicative of random noise rather than a sustained cloud response. The brief positive albedo anomaly at $t=2$ days is arguably consistent with a negative-signed global electric circuit mechanism, though the anomaly is barely larger than the 95% confidence interval and thus difficult to distinguish from noise. The correlation coefficient between the two time series is -0.11 and is not significant at the $p<0.4$ level. We conclude that there is no detectable global albedo response to the CR decrease consistent with a TSI-amplifying mechanism. Furthermore since there is no detectable albedo response larger than the noise in our composite, then it follows with 95% confidence that if a CR–global albedo relationship does exist then a 5% decrease in CRs can, at most, alter global albedo by $\leq 0.0029$. The implications of this upper limit are explored in the discussion below.

The analysis described above was repeated using only the five largest Fd events to test the possibility of a threshold effect (Fig. S5). The analysis was also repeated by partitioning albedos into land and ocean and examining each time series independently (Fig. S6). In both cases there were no statistically significant anomalies at the $p<0.01$ level within 25 days of the Fd minima, and the total number of significant anomalies at the $p<0.05$ level was consistent with the null hypothesis. Furthermore none of the correlation coefficients between the composite albedo and CR anomaly time series were significant at the $p<0.4$ level. Note however that restricting the sample reduces the detectability of albedo responses to CR decreases (see *Laken and Calogovic* [2013] for further discussion).

To test for a latitudinally stratified response, albedo time series for 10 latitude bins were calculated and compared to the CR composite (Fig. S7). There was nothing to suggest a latitudinally stratified albedo response to Fds, with only one incidence of consecutive $p<0.05$ anomalies occurring within 30 days of the Fd minima.

Fig. 3c shows the average global cloud height time series coincident with the Fd events in Fig. 3a. There is a highly significant (albeit wrong signed) negative cloud height anomaly ($p<0.0005$) 17 days after the Fd minima. However this anomaly is likely to be spurious because (i) the anomaly disappears completely when the sample is restricted to the five strongest Fd events (Fig. S9), (ii) it is difficult to imagine a physical mechanism that would cause global cloud height to sharply decrease precisely 17 days after Fd minima for only one day in duration, and (iii) there are no clear altitudinal (Fig. S8) or latitudinal (not shown) responses at 17 days. Assuming the anomaly at 17 days is spurious, the lack of a cloud height response allows us to conclude with 95% confidence that if a CR–cloud top height relationship does exist then a 5% decrease in CRs can, at most, alter global average cloud-top height by 60m. Radiative-convective modeling [*Davies*, 2013] suggests that a 60m change in effective cloud height would cause a 0.4°C change in surface temperature.

## 5. Discussion

Long-term analysis of 13 years of MISR data reveals no statistically significant correlations between the CR flux and global albedo or globally averaged cloud height on monthly or interannual timescales. Additionally there are no statistically significant lagged correlations, and no evidence for any regional correlations.

Epoch superposition of 14 Fd events also reveals no evidence for any albedo or cloud height response to decreases in the CR flux on daily to weekly timescales. Stratifying data by latitude, surface type and restricting analysis to only the largest Fds does not reveal any significant albedo responses inconsistent with the null hypothesis. In this case we can use the null result to constrain the maximum possible influence of CRs on global albedo, and find that if a CR–global albedo relationship does exist then a 5% decrease in the CR flux can, at most, alter global broadband albedo by ≤0.0029.

Given this result we can ask the question: if the CR modulation of global albedo was responsible for recent global warming, would we expect to see a signal in our global albedo composite (Fig. 3b) given the magnitude of the Fds and the noise in our data? The observed increase in global average surface temperature since 1900 is around 0.8°C [*Hansen et al*., 2010]. If we assume a conservative climate sensitivity to global albedo changes of $\lambda = 0.5°\text{C}/(\text{Wm}^{-2})$, then it follows that the necessary secular change in albedo to explain global warming since 1900 is 0.0047. The observed decrease in the CR flux since 1891 is – $5.2 \pm 1.6\%$ (Fig. S10). Thus to explain observed global warming via CR modulation of albedos it is necessary to postulate that a 5% decrease in the CR flux decreases global albedo by around half a percent. Fortuitously the secular decrease in the CR flux since 1891 is equal to the average Fd magnitude in our composite analysis. Thus if CR flux decreases were responsible for recent warming then an albedo signal should be visible in our Fd analysis since the 95% confidence interval extends to only 0.0029. Instead we observe that there is no global albedo response to a 5% decrease in CRs greater than 0.0029 and no hint of any weaker signals imbedded in the noise either. We note that this approach understates the case for expecting to detect an albedo response since (i) the 21-day running mean subtraction tends to reduce the magnitude of Fds by 1-2%, (ii) the cutoff rigidity of the CR monitor used in Fd analysis (R=1.14 GV) is slightly greater than the cutoff rigidity of the CR reconstruction used to determine long term trends (R=0.8 GV), and (iii) since Fds last several days we expect any albedo response to similarly last more than a day, but p-values have been calculated assuming independent daily anomalies (in other words the probability of there being a sustained albedo response of a given magnitude hidden within the noise diminishes as the expected response time increases). Although our conservative value for climate sensitivity arguably overstates the case for detectability, we believe such a choice is justified for two reasons. Firstly if sensitivity were high then the climate system would not yet be in equilibrium and thus using the full sensitivity would be misleading, and secondly, a high sensitivity to radiative forcings contradicts the starting assumption that cosmic rays are responsible for recent warming; if sensitivity were high then 20$^{th}$ century greenhouse gas increases would have caused observed warming contrary to this assumption. It should also be noted that the secular trend in CRs in the last 50 years (in which time 0.6°C of warming has occurred) is 1.4%. This trend is of the wrong sign and 4.4 times too small to explain recent warming given the 95 percentile upper limit on the CR influence from Fig. 3b. The analysis above suggests that CR modulation of albedo is not responsible for the majority of global warming since 1900.

One caveat on these conclusions is that the upper bound of 0.0029 per 5% CR decrease was derived exclusively in the context of short timescale Fds. It is conceivable that CRs influence climate via some unknown mechanism that only acts on longer timescales, and thus wouldn't be apparent during brief Fd episodes. However the majority of CR mechanisms proposed in the literature would be expected to manifest themselves on short timescales since the effects of CRs on atmospheric ionization are immediate, and cloud-formation processes operate on the order of hours to days.

Although both short and long term analysis did not uncover any evidence for spatially localized CR–cloud correlations, local effects cannot be dismissed because the grid size of block-averaged MISR data is large and the sampling errors in regional correlations are too large to tightly constrain the magnitude of local effects.

**Acknowledgements.** We sincerely thank Benjamin A. Laken (Instituto de Astrofísica de Canarias), Abhnil A. Prasad (UNSW) and the two anonymous reviewers for their many helpful comments. We acknowledge the NMDB database founded under the European Union's FP7 program for providing data. Kerguelen neutron monitor data were kindly provided by the French Polar Institute and by Paris Observatory. The original MISR datasets were obtained from the NASA Langley Research Center Atmospheric Science Data Center.

## Figures

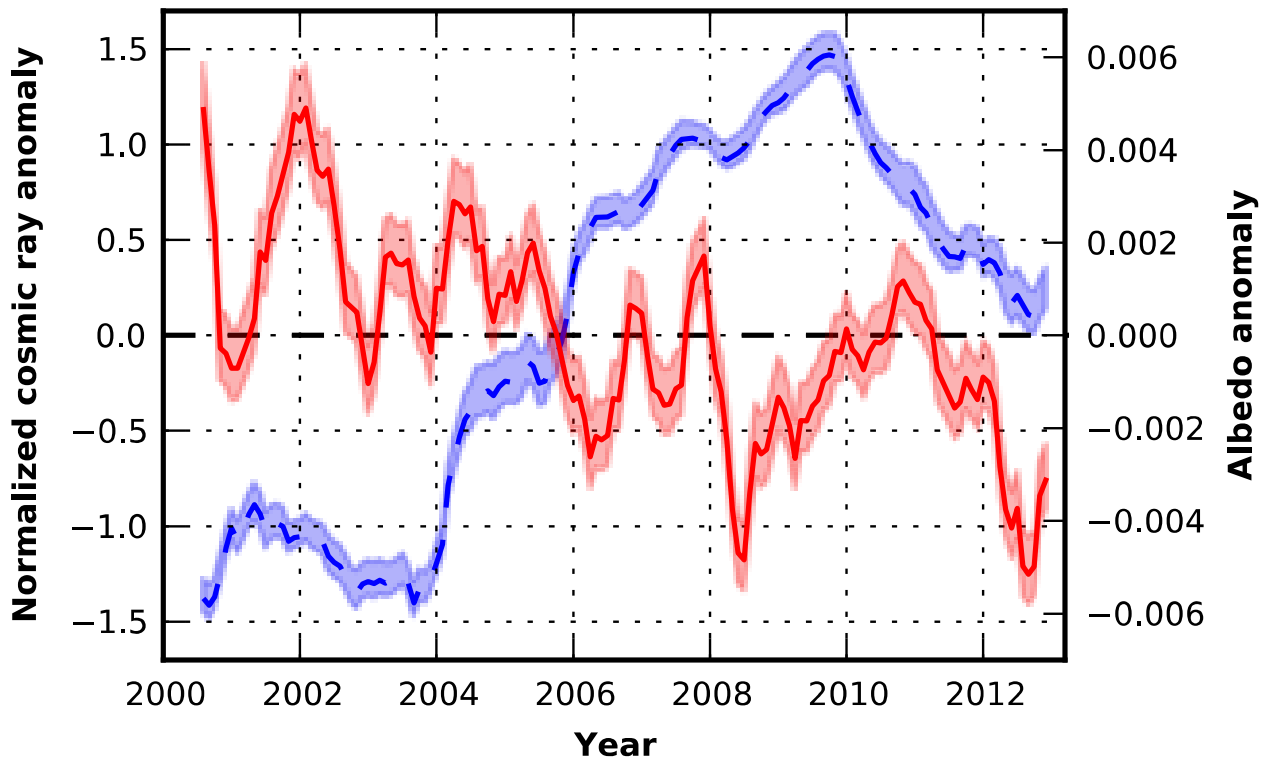

*Figure 1*. Normalized CR anomaly (dashed blue line) and global albedo anomaly (continuous red line) plotted as a function of time. Albedos were retrieved in the green channel (558nm) and both time series were smoothed using a flat 7-month window, though results were robust to changes in the smoothing window and wavelength. Shaded confidence intervals denote ±1 SE.

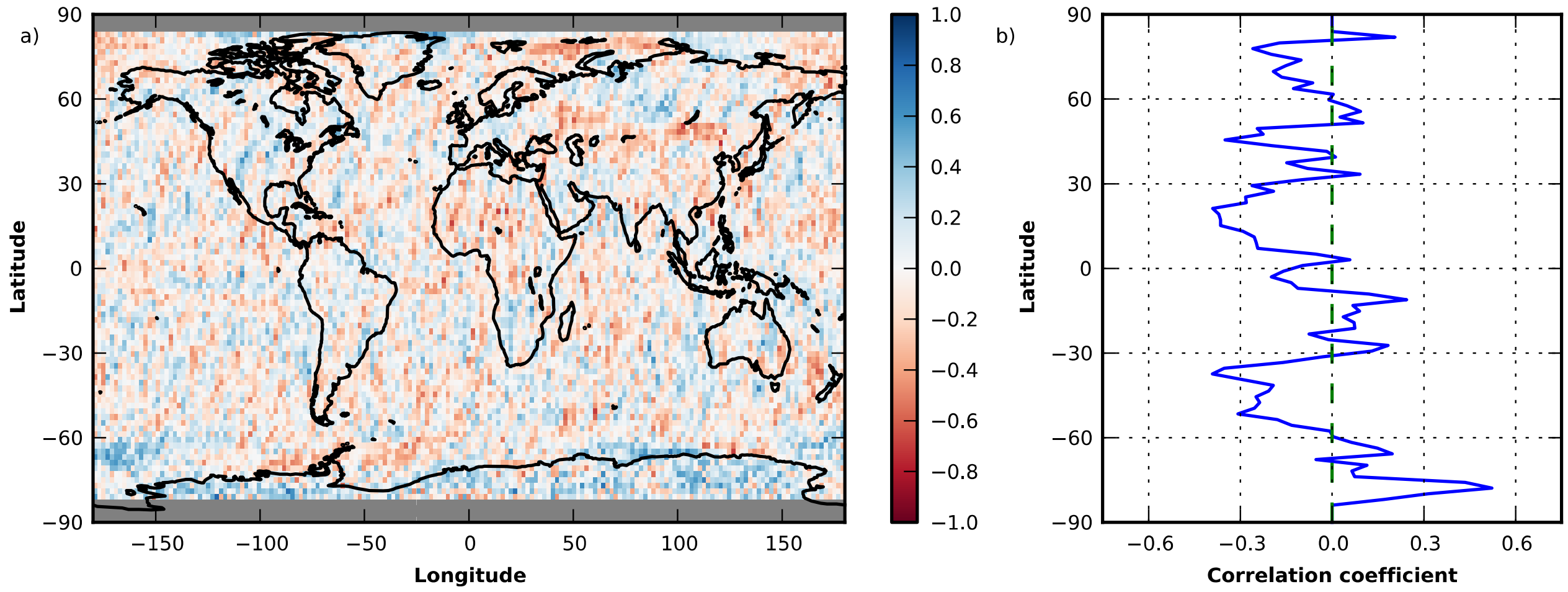

*Figure 2*. (a) Correlation coefficient between albedo and normalized CR anomalies as a function of latitude and longitude (1° by 1° resolution). (b) Correlation coefficient as a function of latitude (1° zonal bands). Albedos were retrieved in the green channel (558nm) and both time series were smoothed using a flat 7-month window, though results were robust to changes in the smoothing window and wavelength. Both time series have been detrended.

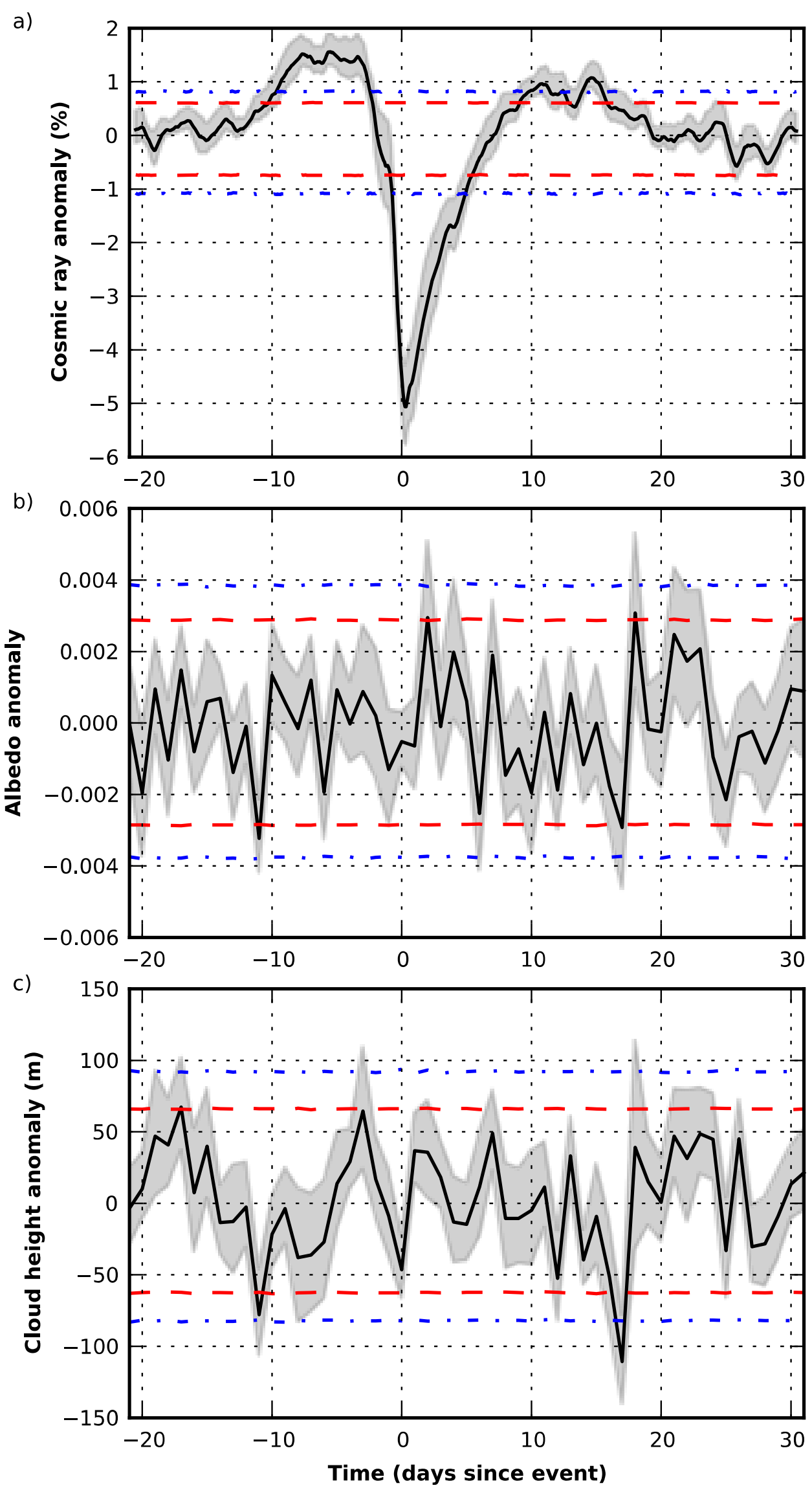

*Figure 3*. Composite time series averaged over all 14 Fd events as a function of days since the Fd minima. Subfigures denote (a) CR anomaly, (b) the corresponding (broadband) global albedo anomaly, and (c) the average global cloud height anomaly. Grey shaded regions denote ±1 SE. Red dashed lines show the limits of the 95% confidence interval and blue dash-dotted lines show the limits of the 99% confidence interval based on 100,000 MC simulations. The average values for confidence intervals are (a) red -0.74 to 0.61 and blue -1.09 to 0.82, (b) red -0.0028 to 0.0029 and blue -0.0038 to 0.0038, and (c) red -62.5 to 66.1 and blue -82.2 to 92.3.

Auxiliary Material 1: Supplementary figures

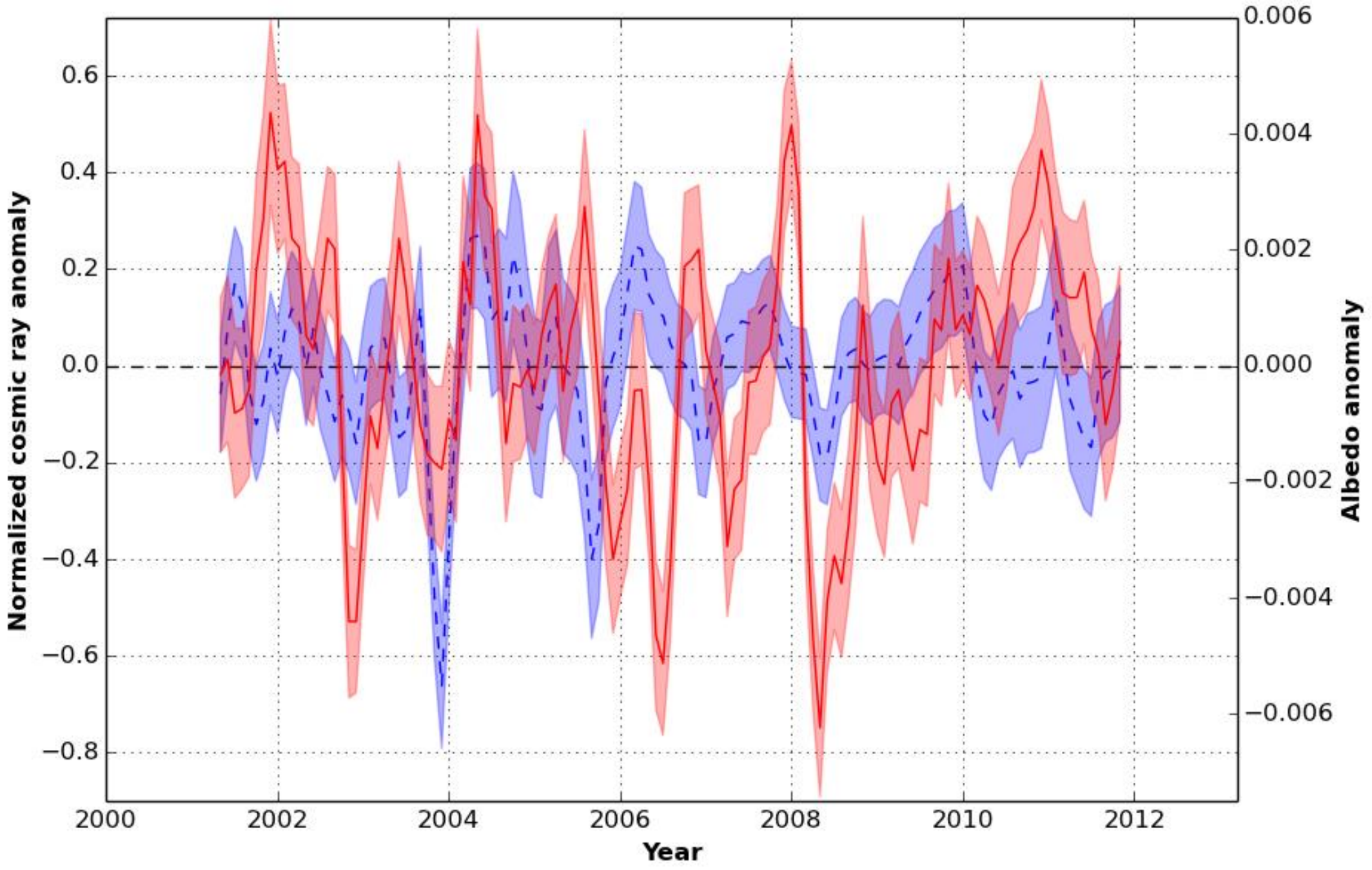

*Figure S1.* Normalized cosmic ray anomaly (dashed blue line) with solar cycle removed as described in main text and detrended global albedo anomaly (continuous red line) plotted as a function of time. Albedos were retrieved in the green channel (558nm) and both time series were smoothed using a flat 3-month window. Shaded confidence intervals denote ±1 standard error.

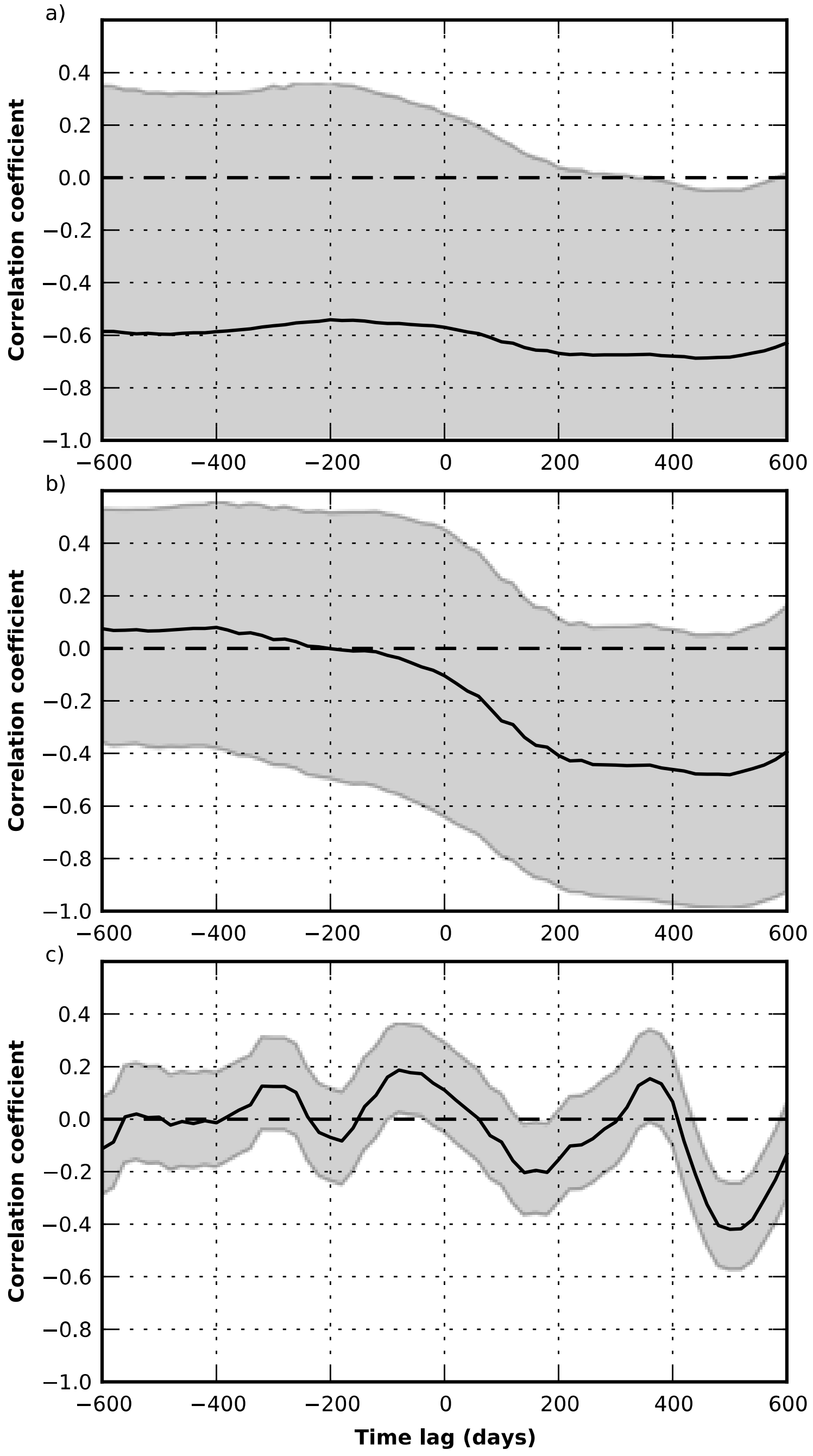

*Figure S2.* Correlation coefficient between global albedo and normalized cosmic ray anomaly as a function of lag for (a) neither time series detrended, (b) both albedo and cosmic rays detrended, and (c) albedos detrended and solar cycle removed from cosmic rays. A positive lag denotes cosmic rays leading albedos (i.e. cosmic rays pushed backwards in time). Grey shading represents ±1 SE. Note that the errors are larger for (a) and (b) due to strong autocorrelation in the cosmic ray time series.

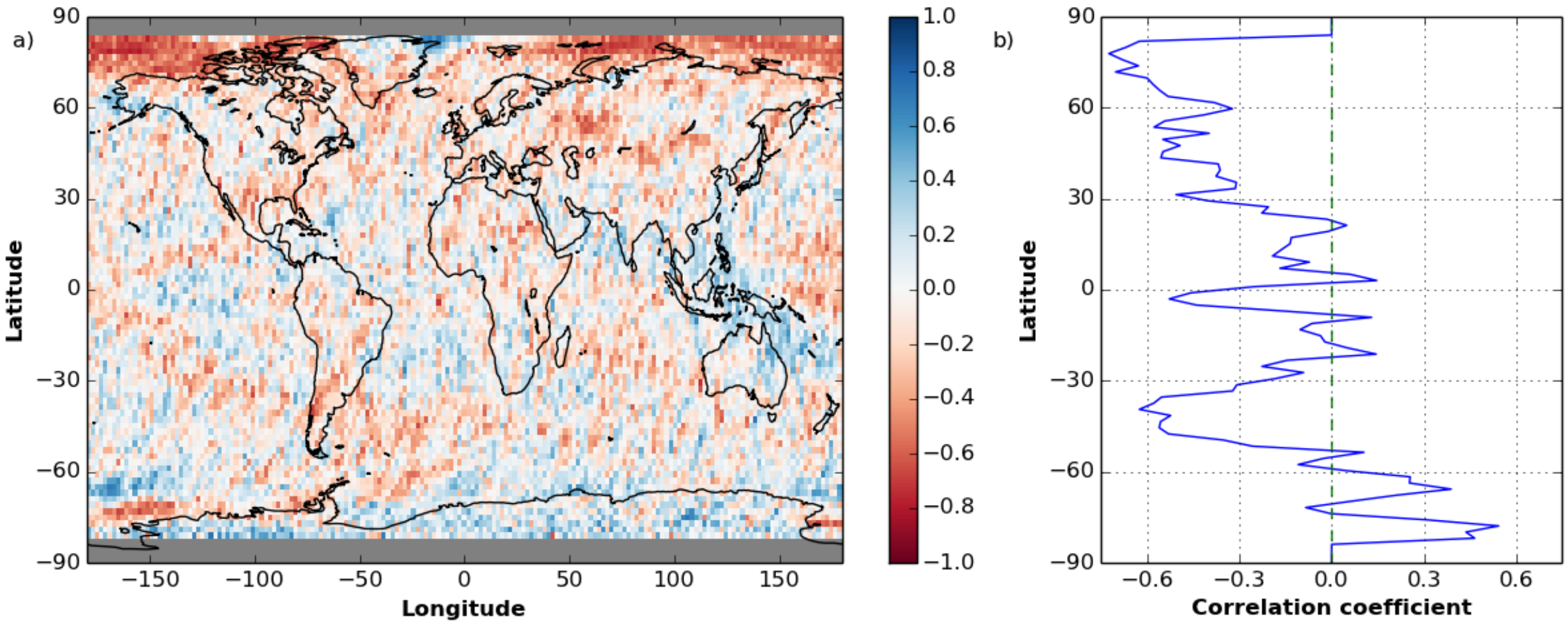

*Figure S3.* (a) Correlation coefficient between albedo and normalized cosmic ray anomalies as a function of latitude and longitude (1° by 1° resolution). (b) Correlation coefficient as a function of latitude (1° zonal bands). Albedos were retrieved in the green channel (558nm) and both time series were smoothed using a flat 7-month window. Neither time series was detrended. The weaker spatial structures in this figure closely resemble the SOI correlations in *Davies and Molloy* [2012].

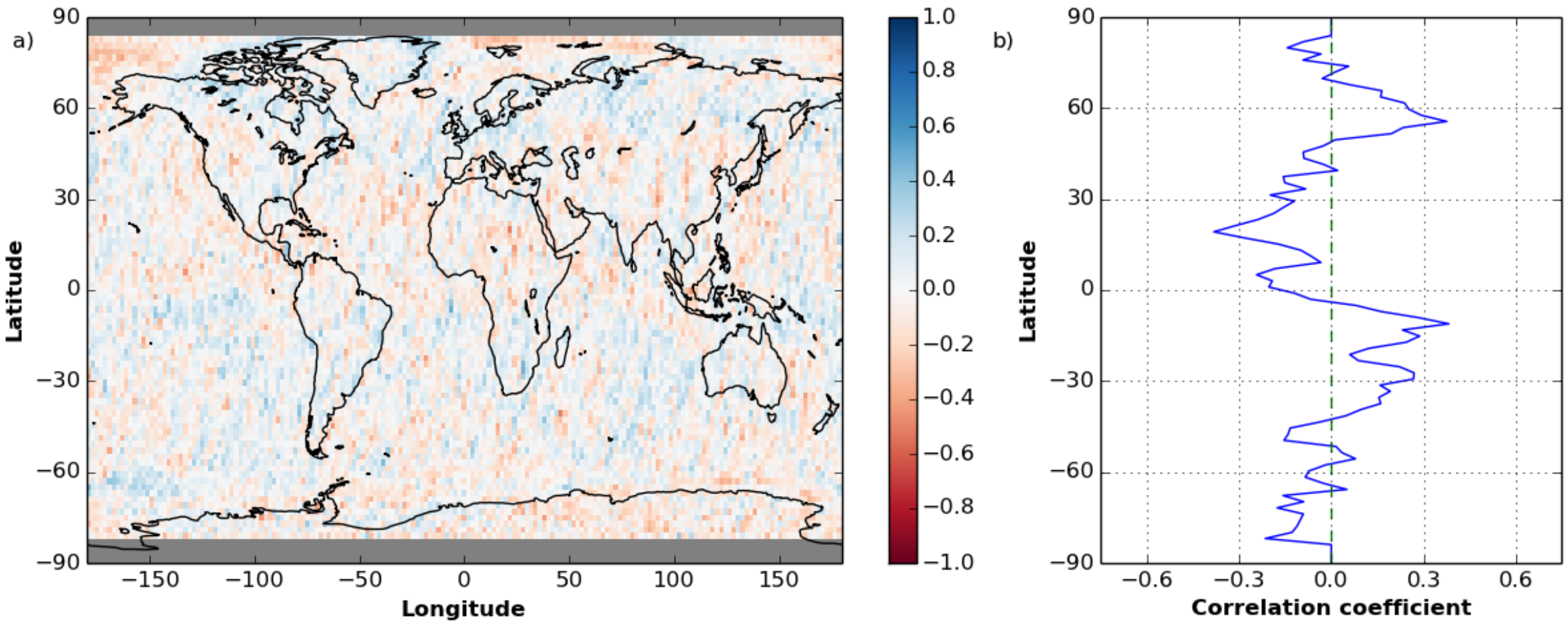

*Figure S4.* (a) Correlation coefficient between albedo and normalized cosmic ray anomalies as a function of latitude and longitude (1° by 1° resolution). (b) Correlation coefficient as a function of latitude (1° zonal bands). Albedos were retrieved in the green channel (558nm) and both time series were smoothed using a flat 3-month window. Albedos have been detrended and solar cycle has been removed from cosmic rays.

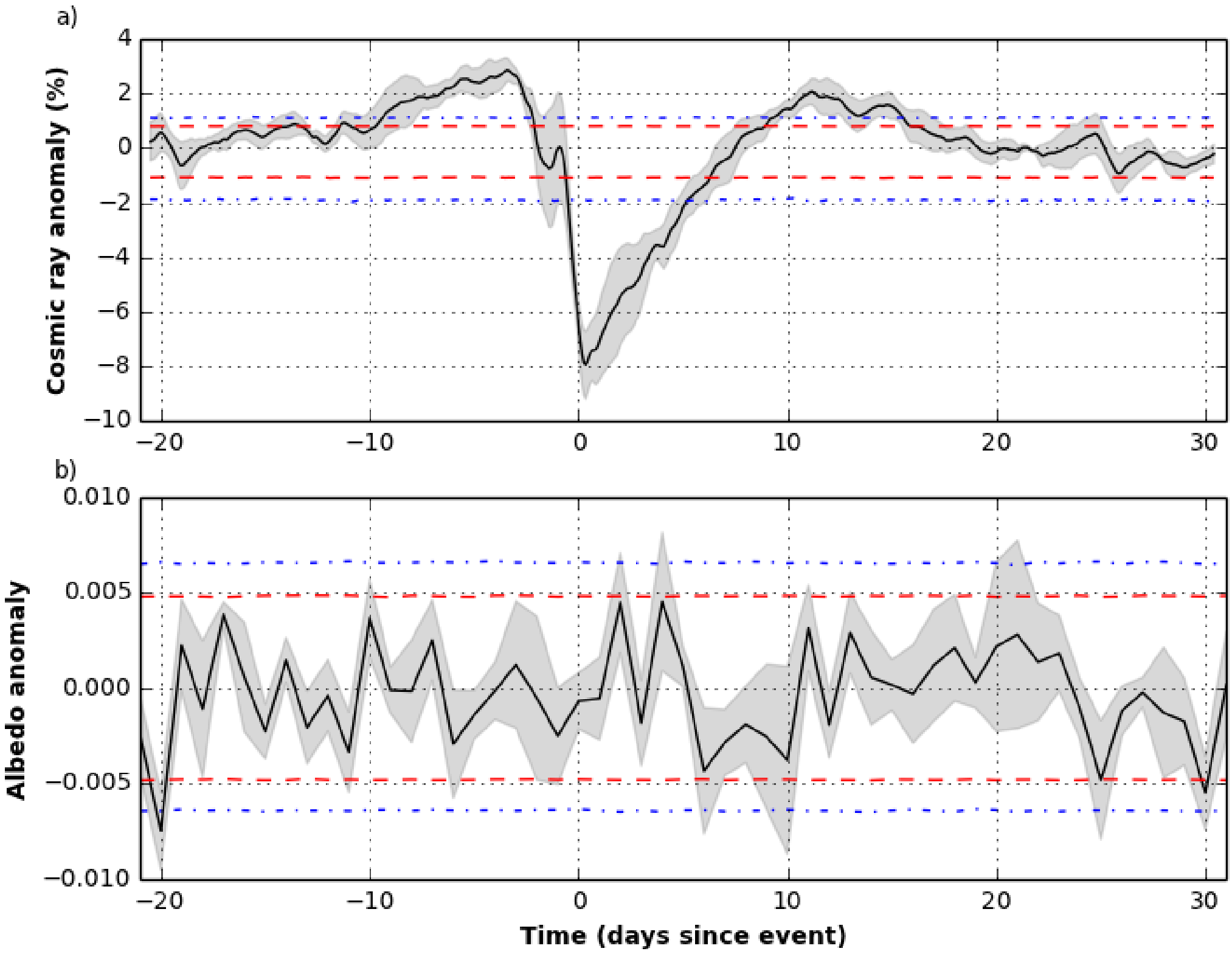

*Figure S5.* Identical to Fig. 3a and Fig. 3b respectively in the main text except analysis restricted to only the five largest Fd events. Grey shaded regions denote ±1 SE. Red dashed lines show the limits of the 95% confidence interval and blue dash-dotted lines show the limits of the 99% confidence interval based on 100,000 MC simulations. The average values for confidence intervals are (a) red -1.09 to 0.81 and blue -1.91 to 1.12, and (b) red -0.0047 to 0.0048 and blue -0.0064 to 0.0066.

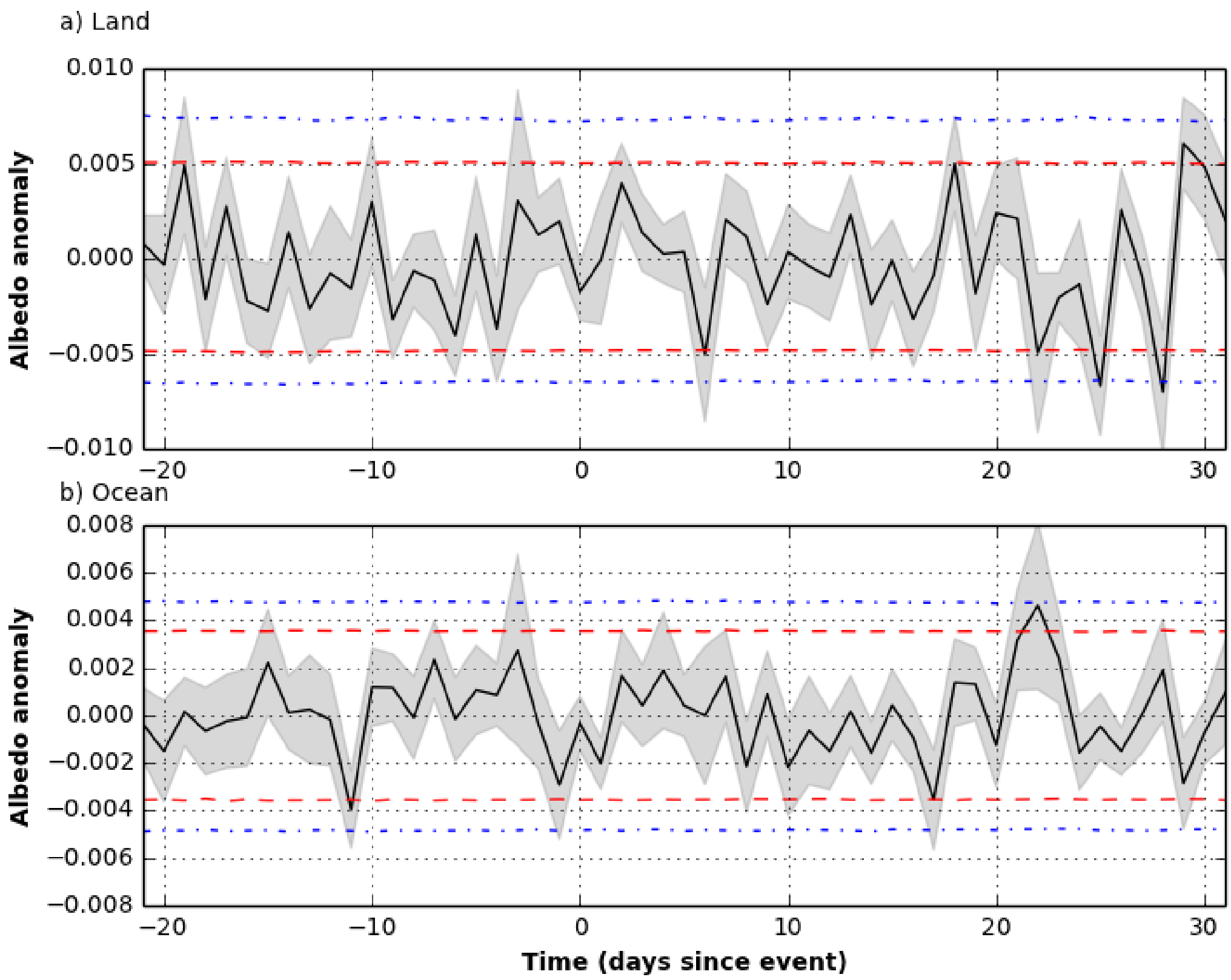

*Figure S6*. Identical to Fig. 3b in the main text except albedo retrievals partitioned into (a) above-land and (b) above-ocean. All 14 Fd events have been used to build these composites. Grey shaded regions denote ±1 SE. Red dashed lines show the limits of the 95% confidence interval and blue dash-dotted lines show the limits of the 99% confidence interval based on 100,000 MC simulations. The average values for confidence intervals are (a) red -0.0048 to 0.0050 and blue -0.0065 to 0.0073, and (b) red -0.0035 to 0.0035 and blue -0.0048 to 0.0048.

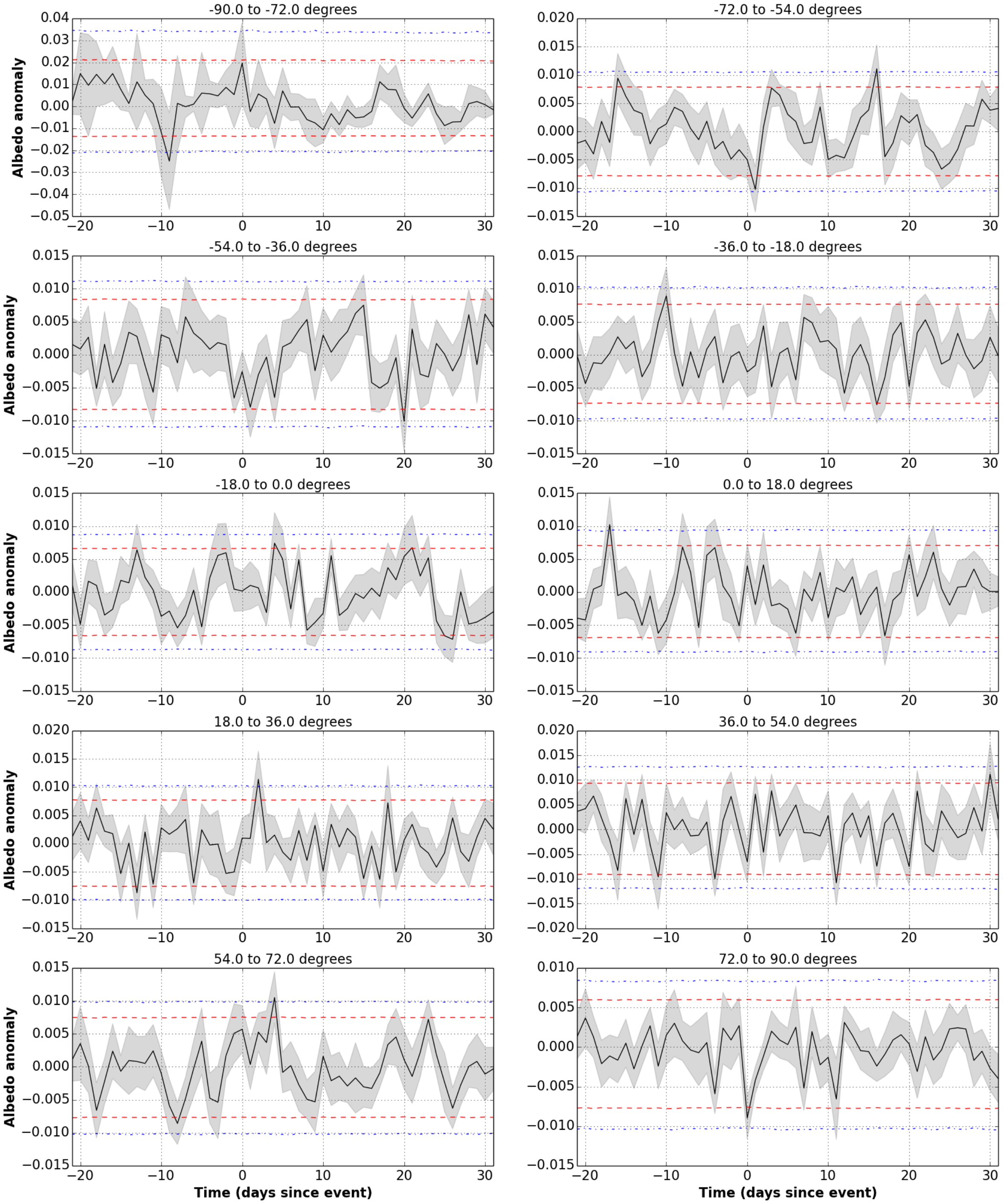

*Figure S7*. Each subfigure shows the composite (broadband) albedo time series for a different latitude band averaged over all 14 Fd events. 4% of the anomalies in all the time series are significant at the $p<0.05$ level which is consistent with the null hypothesis. Grey shaded regions denote ±1 SE. Red dashed lines show the limits of the 95% confidence interval and blue dash-dotted lines show the limits of the 99% confidence interval based on 100,000 MC simulations. The average values of confidence intervals are (red interval then blue interval from negative to positive latitudes): -0.013 to 0.021, -0.021 to 0.034; -0.0078 to 0.0079, -0.011 to 0.011; -0.0083 to 0.0084, -0.011 to 0.011; -0.0074 to 0.0077, -0.0097 to 0.010; -0.0066 to 0.0066, -0.0087 to 0.0088; -0.0069 to 0.0071, -0.0091 to 0.0094; -0.0075 to 0.0077, -0.0099 to 0.010; -0.0091 to 0.0094, -0.012 to 0.013; -0.0077 to 0.0075, -0.010 to 0.0099; -0.0077 to 0.0060, -0.010 to 0.0084.

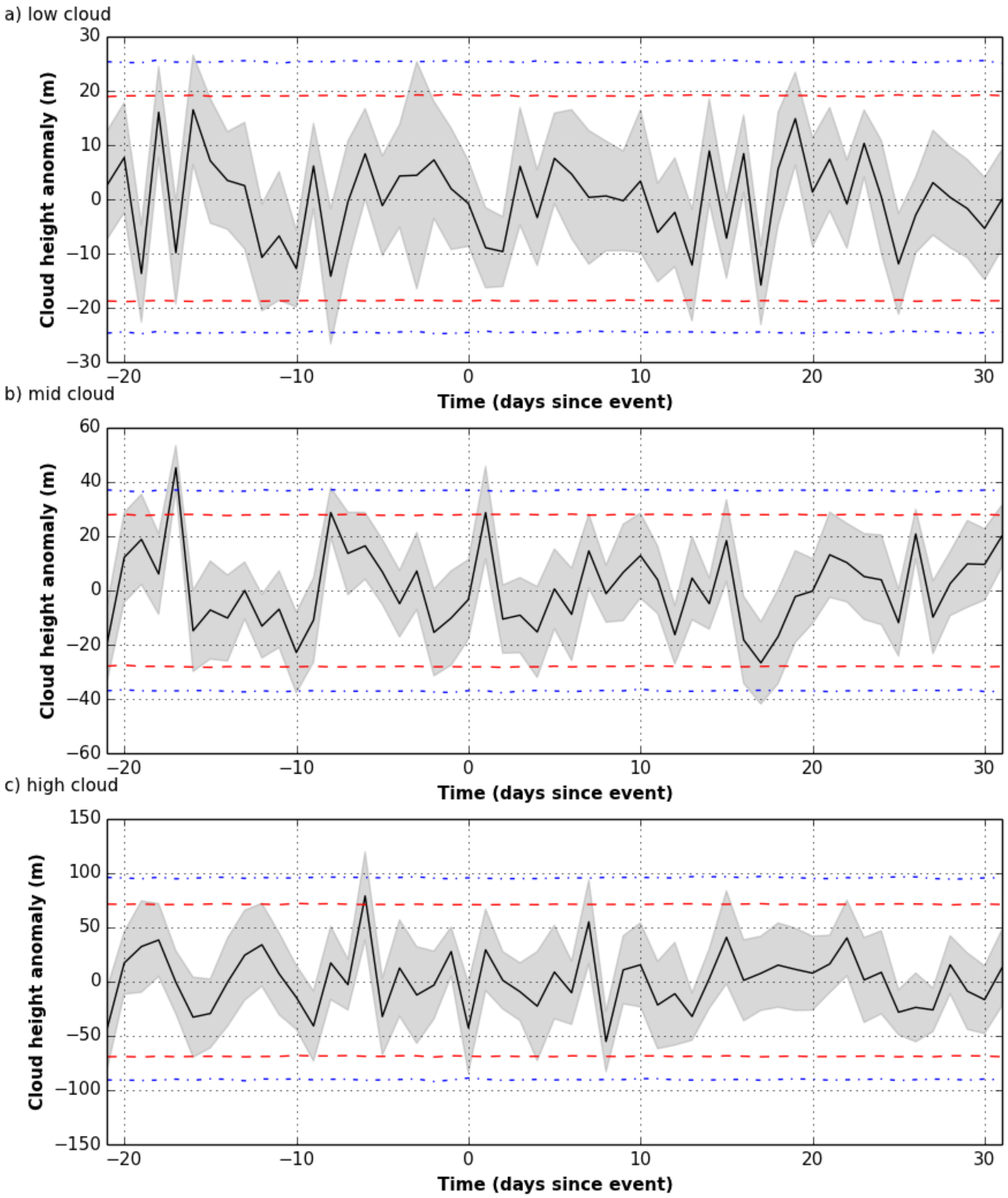

*Figure S8*. Average global cloud height response to 14 Fd events for (a) low, (b) middle and (c) high level cloud. Confidence intervals and standard errors denoted as above. We note that there is no statistically significant anomaly in low, middle or high cloud at t=17 days. The average values for confidence intervals are (a) red -18.6 to 19.1 and blue -24.5 to 25.4, (b) red -28.0 to 27.9 and blue -37.0 to 36.9, and (c) red -68.9 to 71.1 and blue -90.4 to 95.5. *Harrison et al.* [2011] reports a change in the cloud base anomaly distribution when comparing high and low neutron count terciles. It is difficult to directly compare this result to our findings due to uncertainties in how changes in cloud base height translate to cloud top height, and differences in methodology. However we note that Fig. S8a precludes a global low cloud top response of more than 19m per 5% decrease in the cosmic ray flux.

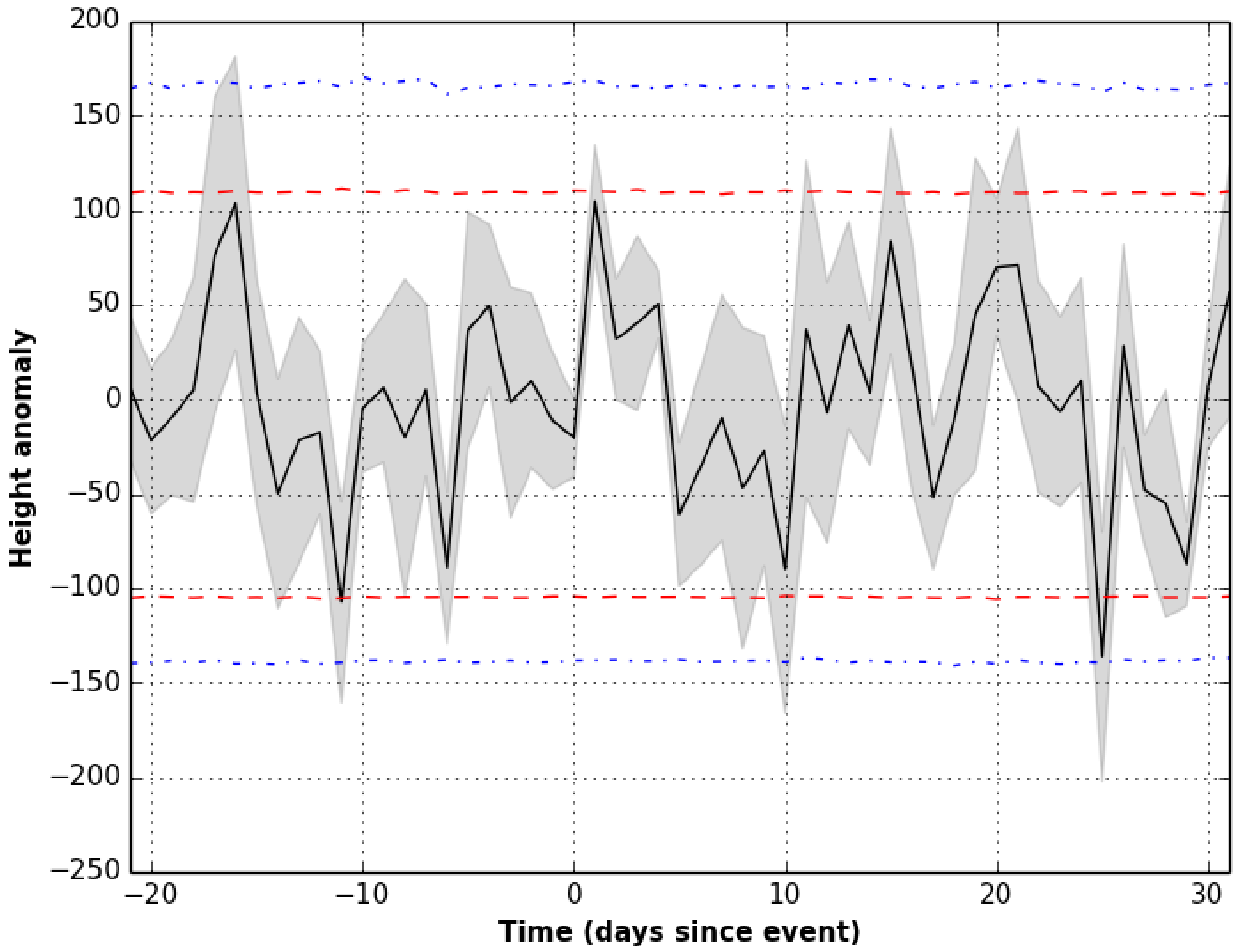

*Figure S9.* Identical to Fig. 3c in the main text except analysis restricted to only the five largest Fd events. Grey shaded regions denote ±1 SE. Red dashed lines show the limits of the 95% confidence interval and blue dash-dotted lines show the limits of the 99% confidence interval based on 100,000 MC simulations. Note how the significant anomaly at t=17 days is now absent. The average values for confidence intervals are red -104.6 to 109.6 and blue -138.5 to 166.4.

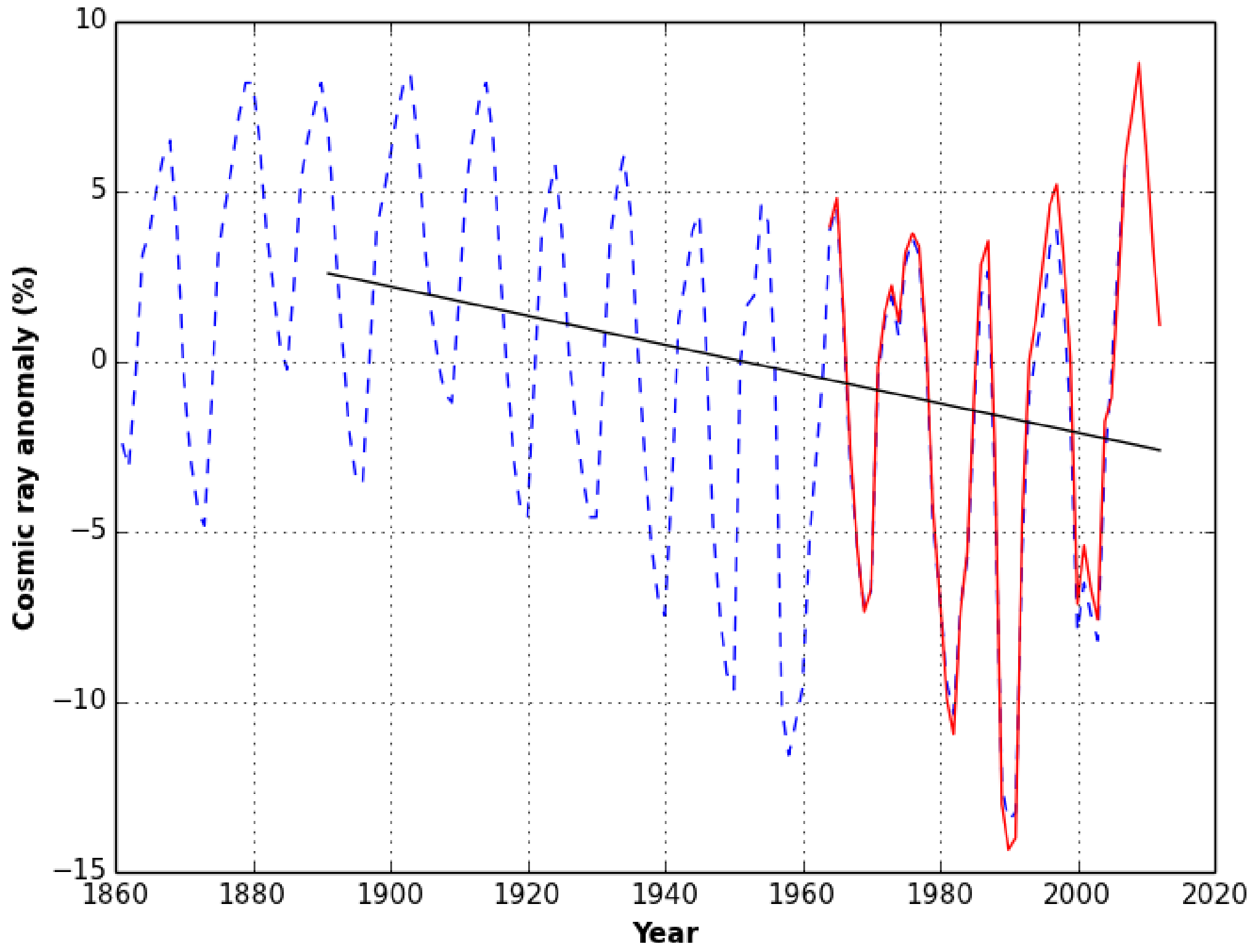

*Figure S10.* Blue dashed line is historical cosmic ray reconstruction from *Usoskin et al.* [2002] for a neutron monitor with rigidity R=0.8 GV, red continuous line is NMBD neutron monitor data from Oulu monitoring station (R=0.8 GV) and the black continuous line shows the secular trend in cosmic rays from 1891 to present (11 complete solar cycles) using the combination of these two time series. The overall change since 1891 is – 5.2 ± 1.6%. The error in the trend was calculated assuming zero autocorrelation, though if autocorrelation is accounted for then (depending on methodology) the trend may not be statistically significant.

Cosmic Ray Data

Long term analysis:

Cosmic ray data were retrieved from the Neutron Monitoring Database (NMDB, www.nmdb.eu). 8 different NMBD cosmic ray monitoring stations were used to build cosmic ray composite time series that represented the global flux for the 13 years of MISR's operation. Each individual time series was anomalized, scaled by its own variance and averaged to produce a 13-year cosmic ray composite as described in main text. Averaging the cosmic ray flux from multiple stations isn't strictly necessary since all stations are very closely correlated with one another, however it does help smooth out minor local variations to better represent the globally averaged surface flux. The stations used were as follows:

Alma-Ata B (R=6.69 GV), AATB
Rome (R=6.27 GV), ROME
IGY Jungfraujoch (R=4.49 GV), JUNG
NM64 Jungfraujoch (R=4.49 GV), JUNG1
Kiel (R=2.36 GV), KIEL
Kerguelen (R=1.14 GV), KERG
Oulu (R=0.81 GV), OULU
Apatity (R=0.65 GV), APTY

Short term analysis:

Only the Kerguelen station (R=1.14 GV) in the French Southern and Antarctic Lands was used in the Fd analysis. This was to ensure the percentage variations in the cosmic ray flux could easily compared to the magnitude of variations on decadal timescales as measured by the Usoskin et. al. (2002) reconstruction (R=0.8 GV). The cosmic ray time series had 2-hourly resolution but was smoothed and interpolated appropriately to match daily sampling in the albedo time series.

14 Forbush decrease events used in epoch superposition analysis:

9/6/2000
18/9/2000
29/11/2000
26/9/2001*
23/3/2002
18/4/2002
31/5/2003
29/10/2003*
15/9/2004
10/11/2004*
12/9/2005*
4/8/2010
18/2/1011
8/3/2012*

* denotes 5 largest events used in sub-sample analysis

The analysis period was chosen to be 53 days; 21 days prior to the minima plus 31 days after the minima plus the minima itself. Arnold (2006) suggests ~7 days as the maximum possible atmospheric response time to CR-induced ionization. The recovery time of Forbush decreases is generally several days. Thus the analysis period was chosen to be several times longer than these events to ensure any signals are clearly distinguishable from background noise (i.e. to provide a clear baseline to contrast with any cosmic ray response).